\documentstyle[aps]{revtex}

\begin{document}
\title{Experimental observation of high field diamagnetic fluctuations in Niobium}
\author{S. Salem-Sugui Jr.$^{1*}$, Mark Friesen$^{2}$, A. D. Alvarenga$^{3}$, F. G.
Gandra$^{4}$, M. M. Doria$^{1}$, and O. F. Schilling$^{5}$}
\address{$^{1}$Instituto de F\'{\i }sica, Universidade Federal do Rio de Janeiro\\
C.P.68528, 21945-970 Rio de Janeiro, RJ, Brasil\\
$^{2}$Department of Materials Science and Engineering, 1500 Engineering\\
Drive, University of Wisconsin, Madison, Wisconsin 53706\\
$^{3}$Centro Brasileiro de Pesquisas Fisicas, Rua Dr. Xavier Sigaud,150,\\
22290-180 Rio de Janeiro, RJ, Brazil\\
$^{4}$Instituto de Fisica, UNICAMP, CP6185, 13083-970 Campinas, SP, Brazil\\
$^{5}$Departamento de F\'{i}sica, Universidade Federal de Santa Catarina,\\
88040-900 Florian\'{o}polis,SC, Brazil.}
\maketitle

\begin{abstract}
We have performed a magnetic study of a bulk metallic sample of Nb with
critical temperature $T_{c}=8.5$~K. Magnetization versus temperature ($M$ 
{\it vs} $T$) data obtained for fixed magnetic fields above 1~kOe show a
superconducting transition which becomes broader as the field is increased.
The data are interpreted in terms of the diamagnetic lowest Landau level
(LLL) fluctuation theory. The scaling analysis gives values of the
superconducting transition temperature $T_{c}(H)$ consistent with $H_{c2}(T)$%
. We search for universal 3D~LLL behavior by comparing scaling results for
Nb and YBaCuO, but obtain no evidence for universality.

\end{abstract}

High field diamagnetic fluctuations are predicted to occur in
superconductors in a strong magnetic field, $H$. This constrains the paired
quasi-particles to remain in the lowest Landau level (LLL), reducing their
effective dimensionality \cite{a1,a2}. The width of the region around the
superconducting temperature transition, $T_{c}(H)$, where the high field
fluctuations occur is given by the field dependent Ginzburg criterion $%
G(H)=(8\pi \kappa ^{2}k_{B}T_{c}H/\phi _{0}\xi _{c}H_{c2}^{2})^{2/3}$ \cite
{a3}, where $\kappa $ is the Ginzburg-Landau (GL) parameter, $\phi _{0}$ is
the quantum flux, $\xi _{c}$ is the c-axes coherence length at zero
temperature, and $H_{c2}$ is the upper critical field at zero temperature
and for fields applied along the c-axes direction. One important effect of
the LLL fluctuations is to produce a rounding of various data curves around $%
T_{c}(H)$ \cite{a4,a5}. Consequently, the superconducting transition appears
continuous, rather than distinct, as expected for a second order transition $%
T_{c}(H)$ \cite{a6}. High $T_{c}$ superconductors (HTSC), with their small $%
\xi $, high $\kappa $, and high critical temperatures, $T_{c}$, display a
broad fluctuation region. The LLL fluctuation theory has been invoked to
explain the nature of the broad ``fan''-shaped transition observed in
high-field magnetic measurements in YBaCuO \cite{a6}. Predictions of the LLL
theory include scaling laws for various physical quantities \cite{a4,a5,a5a}%
. For magnetization in particular, the scaling predicts that $M$ {\it vs} $T$
data obtained at different fields, $H$, should collapse onto a single curve
when the variable $M/(TH)^{(D-1)/D}$ is plotted against $%
(T-T_{c}(H))/(TH)^{(D-1)/D}$. Here, $T_{c}(H)$ becomes a fitting parameter,
and $D$ is the dimensionality of the system. This scaling law has been used
to identify LLL fluctuations in a given material, and determine its
dimensionality \cite{a5,a6,a7,a8,a9,a10,a10a,a11}. An important check of the
scaling is that it should provide reasonable values of $T_{c}(H)$ \cite
{a6,a7}. For lower-dimensional or layered materials, the LLL analysis also
helps to explain the crossing points observed in $M(T)$ curves, \cite
{a8,a9,a10,a11,a12}.

Despite the relevance of LLL fluctuations for HTSC there are few comparable
studies in conventional superconductors. Niobium is one of the most studied
type II superconductors. Observations of an irreversibility line and a broad
high-field reversible region in Nb \cite{a14}, similar to HTSC, have renewed
interest in this element. Extensive studies of fluctuations have been
performed in Nb at low fields \cite{a15}. For pure samples, high field
diamagnetic fluctuations were observed in specific heat measurements for $%
H=4 $~kOe, but only over a narrow temperature range, of order 40~mK \cite
{a13}. It is therefore a matter of interest to gather further evidence for
high field diamagnetic fluctuations in Nb. Our work is motivated by the
possibility that a Nb sample with an elevated value of $\kappa >1$ could
show enhanced high field fluctuation effects.

We address this issue by performing magnetization measurements as a function
of field and temperature in an impure but homogeneous Nb sample. The results
show that $M(T)$ curves obtained for different fields follow the 3D LLL
scaling. The Nb curves are somewhat more separated than comparable curves in
HTSC, where LLL scaling has been applied \cite{a6,a7,a8,a9,a10,a11,a12}. The
crossing point behavior sometimes observed in HTSC is not here observed for
Nb, as consistent with its expected 3D nature.

The Niobium sample adopted in the present study is an ellipsoid with axes $%
2r_{1}$ = 4.7 mm and $2r_{2}$ = 5.3 mm, mass = 0.6487 g, and $T_{c}=8.5$~K
with $\Delta T=0.3$~K determined at low fields. The sample was manufactured
in an arc melt furnace from 99.9 \% Nb wire. X-ray diffraction shows the
metallic Nb phase. Magnetization data were always taken after cooling the
sample in zero field. A commercial Quantum Design SQUID magnetometer (7T)
with 3 cm scans was utilized in the measurements. Hysteresis loops of $M$ 
{\it vs} $H$ were obtained at fixed temperatures with values running from 3
K to 6 K. Isofield magnetization curves, $M$ {\it vs} $T$, were also
obtained, for fixed applied magnetic fields with values running from 1 kOe
to 10 kOe. The value of $T_{c}$ = 8.5 K, lower than 9.2 K found for pure Nb 
\cite{a17}, indicates the possible existence of magnetic impurities.
Neverthless, the corresponding magnetic signal is very small and could not
be resolved from the background magnetization. The corrected background
magnetization is found to be field dependent but not temperature dependent,
as expected for Pauli paramagnetism in the studied temperature range. The
demagnetization factor of the sample is obtained from the Meissner region in
hysteresis magnetization curves $M$ {\it vs} $H$ and is very close to 1/3 as
expected for a sphere. A previous study performed in our sample \cite{a18}
determined the value of the Ginzburg parameter $\kappa =4$. Since the
temperature width of the high-field fluctuations can be expressed as $\Delta
T_{{\rm fluct}}\simeq G(H)T_{c}\propto \kappa ^{4/3}$, large values of $%
\kappa $ can be expected to enhance fluctuations effects. High purity Nb,
with $T_{c}=9.2$~K, has $\kappa \simeq 1$ \cite{a17}.

Figure 1 shows zero-field cooled $M$ {\it vs} $H$ hysteresis curves. The
inset shows the superconducting transition $T_{c}$. The entire transition
occurs within a temperature window of less than 0.3 K. However the step,
which accounts for 80\% of the transition, has a 80~mK width, between 8.42
and 8.5~K. Most of the $M(H)$ curves in Fig. 1 display a pronounced peak
effect occurring near the irreversible field $H_{{\rm irr}}$, reminiscent of
HTSC.

Figure 2 shows a detail of the reversible region of the hysteresis $M$ {\it %
vs} $H$ curve obtained at 3.5 K. The reversible diamagnetic region extends
more than 2 kOe above $H_{{\rm irr}}$. Figure 2 shows the standard linear
extrapolation of the reversible magnetization down to M = 0, used to
estimate $H_{c2}$ \cite{a19}. The linear extrapolation is clearly an
approximation, since there is a large region of the curve which deviates
from the linear behavior as the magnetization approaches zero. To this level
of accuracy, $H_{c2}$ is located 1.2 kOe above $H_{{\rm irr}}$. The inset of
Fig. 2 shows a detail of the $M$ {\it vs} $H$ curve obtained at 6 K, for
which the reversible diamagnetic region extends only approximately 300 Oe
above $H_{{\rm irr}}$. The results for $H_{c2}$ and $H_{{\rm irr}}$ obtained
from Fig. 1 are plotted in Fig. 3.

Figure 4 shows zero field cooled $M(T)$ curves obtained for fields $H$ = 10,
9, 8, 7, 6, 5, 3, 2, and 1 kOe. For $H$ = 10, 9, and 8 kOe the field cooled
data are also plotted to show the extent of the reversible regions at high
fields. (We mention that values of $(T_{{\rm irr}},H)$ obtained from $M$ 
{\it vs} $T$ curves agree with values of $(T,H_{{\rm irr}})$ obtained from $%
M $ {\it vs} $H$ curves.) The temperature interval between magnetization
data in each curve is 50 mK. The rounding of the curves at high fields makes
it difficult to extract a value for the critical temperature $T_{c}(H)$. To
illustrate the rounding effect, we mark the transition temperatures, as
interpolated from Fig. 3, with arrows for the $H$ = 1, 2 and 10 kOe curves.
By visual inspection of Fig. 4, it is possible to see a sharp transition for 
$H$ = 1 kOe, which evolves into a broader transition as the field increases.

We now consider possible explanations for the rounded curves in Figure 4. In
Ref.~[17], high field diamagnetic fluctuations were observed for Nb at $H$ =
4 kOe, suggesting that the rounding above 5 kOe in our data may also be due
to LLL-type fluctuations. An alternative explanation for the broadening
could be surface superconductivity, which may occur above $T_{c}(H)$.
However, the surface superconductivity signal for a sphere is expected to be
rather small \cite{a20} as confirmed for our sample in the $M$ {\it vs} $T$
curve for $H$ =1 kOe, where no diamagnetic signal was detected above $%
T_{c}(H)$. Additionally, rounding effects in the magnetization due to
surface superconductivity would only occur above $T_{c}(H)$ \cite{a20} while
in Figure 4 for $H$ = 10 kOe the rounding also occurs below $T_{c}(H)$.
Based on these observations, and the fact that the data follow the 3D~LLL
scaling, we disregard the influence of surface superconductivity in the
present measurements. Another possible explanation for the rounding in $M$ 
{\it vs} $T$ curves is sample inhomogeneity due to the impurities. We
believe that our sample is homogeneous based on three points: (1) the sample
shows only one superconducting transition at low fields which is relatively
sharp, (2) there is no rounding above the transition for the $H=1$~kOe $M$ 
{\it vs} $T$ curve, and (3) the data follow the 3D~LLL scaling.

We now perform a Lowest Landau Level scaling analysis on the high field data
of Fig. 4, with $D=3$, as appropriate for Nb. $M/(TH)^{2/3}$ is plotted {\it %
vs} $(T-T_{c}(H))/(TH)^{2/3}$. The transition temperature, $T_{c}(H)$,
becomes a fitting parameter, chosen to make the data collapse onto a single
curve. The resulting values are plotted in Fig. 3, with a significant
correspondence to the values of $H_{c2}(T)$ estimated from the isothermic $M$
{\it vs} $H$ curves. The results of the scaling analysis are presented in
Figure 5. The inset shows the unscaled data while the collapsed data is
shown in the main figure. Note that only high-field data in the field range $%
H$ = 5 - 10 kOe have been scaled. The arrows in the inset identify the
scaling values of $T_{c}(H)$, evidencing the breadth of the LLL fluctuation
range. A visual inspection of this inset suggests that the LLL fluctuations
occur over a large region in the 9 and 10 kOe curves, including data below $%
T_{c}(H)$. However, for 7, 6 and 5 kOe, the LLL scaling seems appropriate
only above $T_{c}(H)$. We may compare the observed temperature width of the
scaling range to its expected width, $\Delta T_{{\rm fluct}} \simeq
G(H)T_{c}(H)$. $\Delta T_{{\rm fluct}}$ is expected to grow with field as $%
H^{2/3}$. We observe the predicted $H^{2/3}$ behavior for H = 10, 9 and 8
kOe , but not for $H$ = 7, 6, and 5 kOe. It is possible that the high-field
conditions, namely, the paired quasi-particles lying in the lowest Landau
level, are not fully met for the latter data.

The recent interest in LLL scaling has been stimulated by high temperature
superconductors, which exhibit extensive high field fluctuations \cite
{a4,a5,a6,a5a,a7,a8,a9,a10,a10a,a11,a12}. It is of fundamental interest to
ask whether scaling in low and high temperature superconductors could
exhibit universal behavior. To investigate this point we obtain
magnetization measurements from the sample of YBaCuO (mass = 1.7 mg, $%
T_{c}=91.7$~K) used in Ref.~[11]. Although YBaCuO is known to be more
anisotropic than Nb, it is still expected to behave three dimensionally,
justifying the use of 3D-LLL scaling. It is possible that the 3D scaling
form could break down away from $T\simeq T_{c}(H)$, owing to the layered
nature of the material \cite{a5a}. (This is the same regime where the simple
scaling forms used in this work become inaccurate, due to the more
complicated nature of LLL scaling theory in 3D \cite{a5a}.) We perform
simultaneous scaling of YBaCuO and Nb in analogy to Fig.~5, however we now
incorporate sample-dependent scaling factors along both the $x$ and $y$
axes. Specifically, the YBaCuO data has been multiplied by the relative
factors of 1.5 and 1/150, along the $x$ and $y$ axes, respectively. The
results are shown in Fig.~6 . The collapse of data from the two different
samples is good, and may give support to the notion of universality.

We can take this comparison a step further by analyzing the collapse in Fig.
6. Because of the different geometries of the two samples, it is not
possible to make a direct comparison of the magnetizations ($y$-axis).
However, Ullah and Dorsey \cite{a5} provide an analytical expression for the
full 3D-LLL temperature scaling variable along the $x$ axis: $x=(\gamma
/(2\kappa ^{2}-1)TH)^{2/3}(\partial H_{c2}/\partial T)(T-T_{c}(H))$, where
the anisotropy parameter is given by $\gamma =\xi _{c}(0)/\xi _{ab}(0)$. The
ratio of the respective YBCO and Nb scaling factors, $r=[(\gamma /(2\kappa
^{2}-1))^{2/3}\partial H_{c2}/\partial T]_{{\rm YBaCuO}}/[(\gamma /(2\kappa
^{2}-1))^{2/3}\partial H_{c2}/\partial T]_{{\rm Nb}}$, can therefore be used
to rescale the YBCO data to make it collapse onto the Nb data. Using typical
values of $\gamma =0.2$, $\kappa =55$, and $\partial H_{c2}/\partial T=15$%
~kOe/K for YBaCuO, and $\gamma =1$, $\kappa =4$, and $\partial
H_{c2}/\partial T=1.5$~kOe/K for our Nb sample, we calculate $r=0.1$, which
is more than an order of magnitude different than the measured value, $r=1.5$%
. Thus, independently, the Nb and YBCO samples appear to scale as expected.
A naive universal collapse of the data sets appears outwardly successful.
However the detailed scaling results cannot be reconciled with Ref.~%
\onlinecite{a5}.

There are two possible explanations for these mixed results, which do not
necessarily invalidate the observed 3D-LLL scalings: (1) The temperature
range of 3D~LLL scaling could be so narrow for YBCO, due to its layered
nature, that the observed 3D~LLL collapse is either rendered invalid or
belongs to a different universality class. (2) The theory of Ullah and
Dorsey \cite{a5}, employed here to determine $r$, may not be applicable to
isotropic superconductors such as Nb, because of its origins in the layered
Lawrence-Doniach model. The expression we use for $r$ would therefore be
inaccurate, although the basic scaling theory (ignoring scaling factors) is
expected to be independently valid for the two types of superconductors.

There is also a third possible explanation for the failure of universal
scaling. It has been shown in Ref.~\onlinecite{a10a} that 3D~LLL scaling in
YBaCuO is obeyed better for data in the high field region ($H>5$~T) than in
the 1 - 5 T field region. Unfortunately we do not have high field data for
YBaCuO to test this possibility. The inset of Fig. 6 shows YBaCuO unscaled
data used in the main figure. The arrows identify the scaling values of $%
T_{c}(H)$, as in Fig.~5. Comparison of Figs.~5 an 6 shows that LLL scaling
in Nb is all the more spectacular due to the wide separation of the unscaled
data.

Finally, we point out that the $M$ {\it vs} $T$ curves of Fig. 4 exhibit no
crossing points. In HTSC the $M$ {\it vs} $T$ curves for which LLL have been
successfully applied show crossing points \cite{a6,a5a,a7,a8,a9,a10,a11,a12}
which are related to the layered structure (2D character) of those systems.
We also mention that we tried to scale our Niobium data using a 3D~$XY$
critical scaling theory \cite{a21} without success. The 3D~$XY$ theory has
been used to explain fluctuations of HTSC in the reversible regime \cite
{a21,a10a}.

Work partially supported by CNPq, and FAPESP.

* Corresponding author; e-mail: said@ if.ufrj.br

\begin{figure}[]
\caption{$M$ {\it vs} $H$ curves obtained for fixed temperatures ranging
from 3 K to 6 K. Inset shows the superconducting transition for an applied
magnetic field of 5 Oe.}
\end{figure}

\begin{figure}[tbp]
\caption{ Detail of the $M$ {\it vs} $H$ curve obtained at 3.5 K for fields
in the region of $H_{c2}$. Inset shows detail of the $M$ {\it vs} $H$ curve
obtained at 6 K in the region of $H_{c2}$.}
\end{figure}
\begin{figure}[tbp]
\caption{ Magnetic phase diagram for our Nb sample. The plotted values of $%
H_{{\rm irr}}$ and $H_{c2}$ are determined as in Fig.~2. The values of $%
T_{c}(H)$ are obtained from the 3D~LLL scaling analysis.}
\end{figure}

\begin{figure}[tbp]
\caption{ Zero-field-cooled $M$ {\it vs} $T$ curves obtained at fixed
magnetic fields. From left to right, curves were obtained for $H$ = 1, 2, 3,
5, 6, 7, 8, 9, and 10 kOe. For $H$ = 8, 9, and 10 kOe the corresponding
field-cooled data are also plotted.}
\end{figure}

\begin{figure}[]
\caption{ Magnetization data for Nb are plotted after performing a 3D Lowest
Landau Level scaling analysis. The inset shows only the unscaled data
collapsed in the main figure.}
\end{figure}

\begin{figure}[]
\caption{ Simultaneous 3D~LLL scaling of Nb and YBaCuO data. The $x$ and $y$%
-axes of YBaCuO data are multiplied by 1.5 and 1/150 respectively. Open
symbols are used for Nb data, and filled symbols for YBaCuO data. The inset
shows the unscaled $M$ {\it vs} $T$ data of YBaCuO used in the main figure.}
\end{figure}

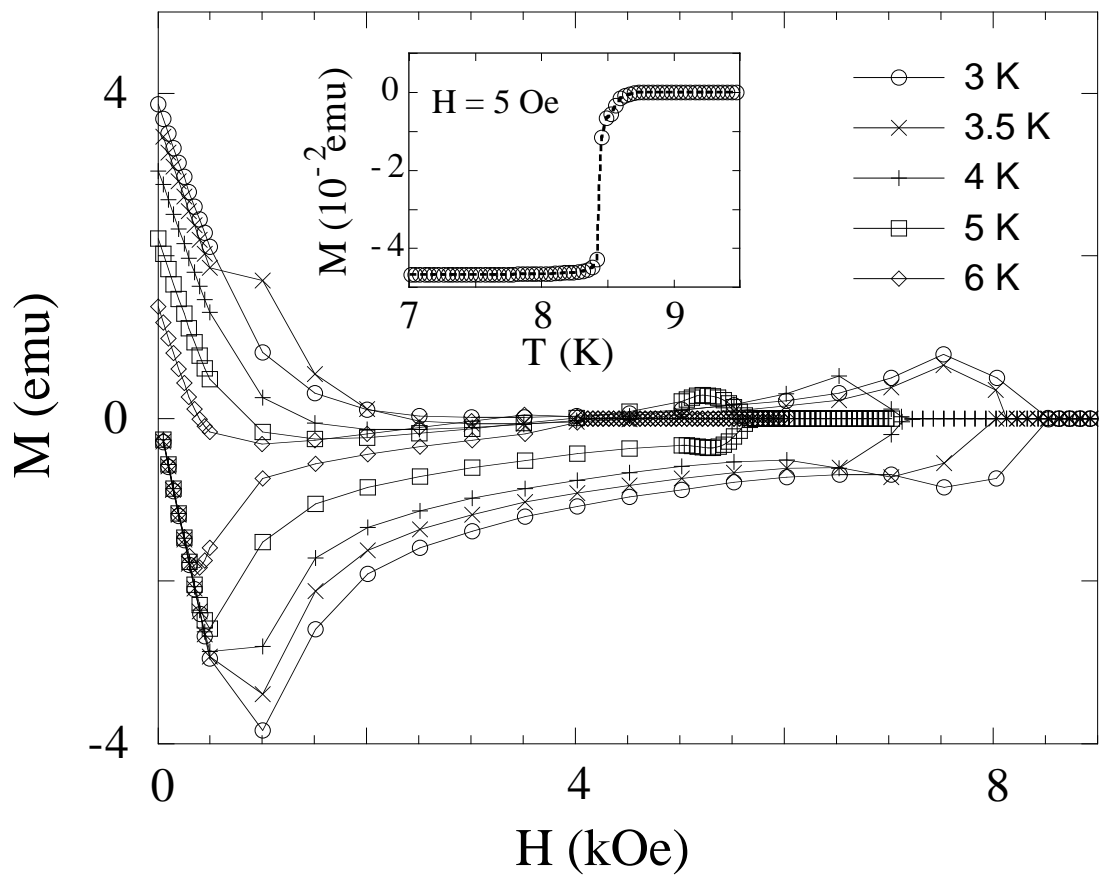
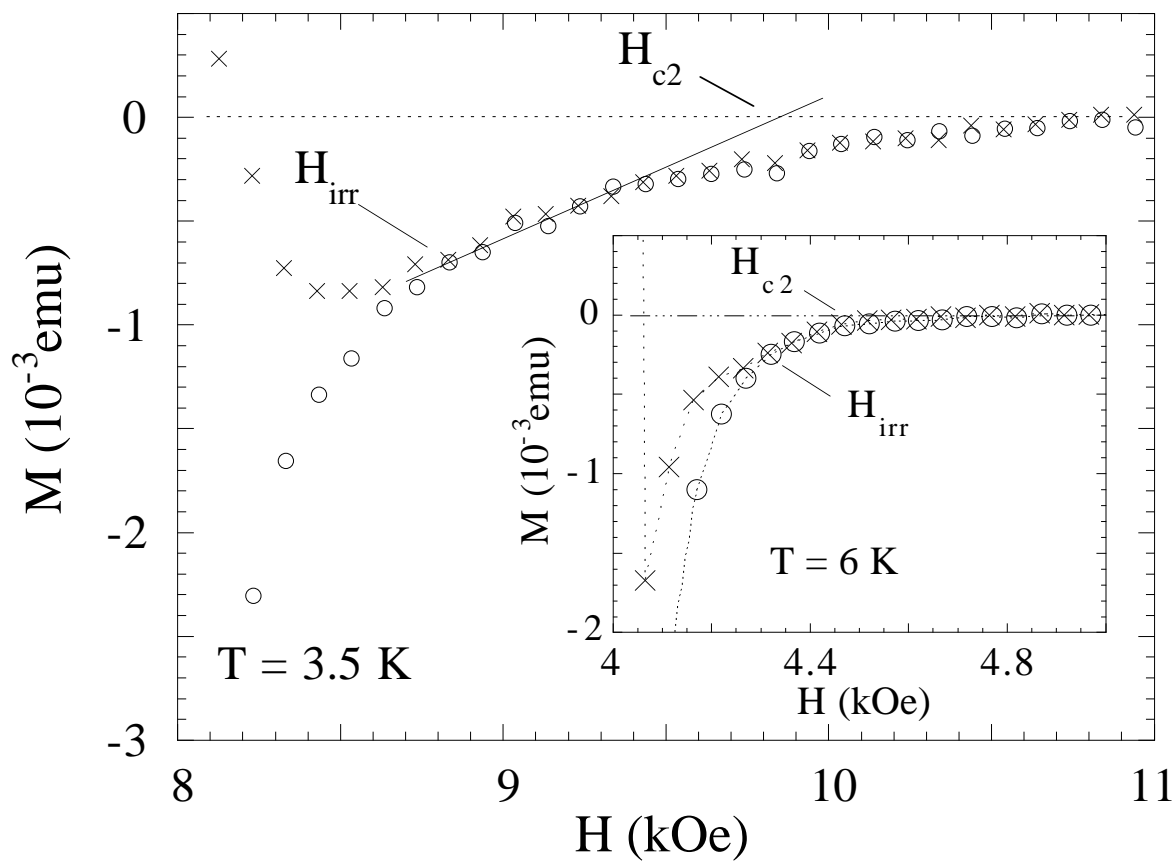
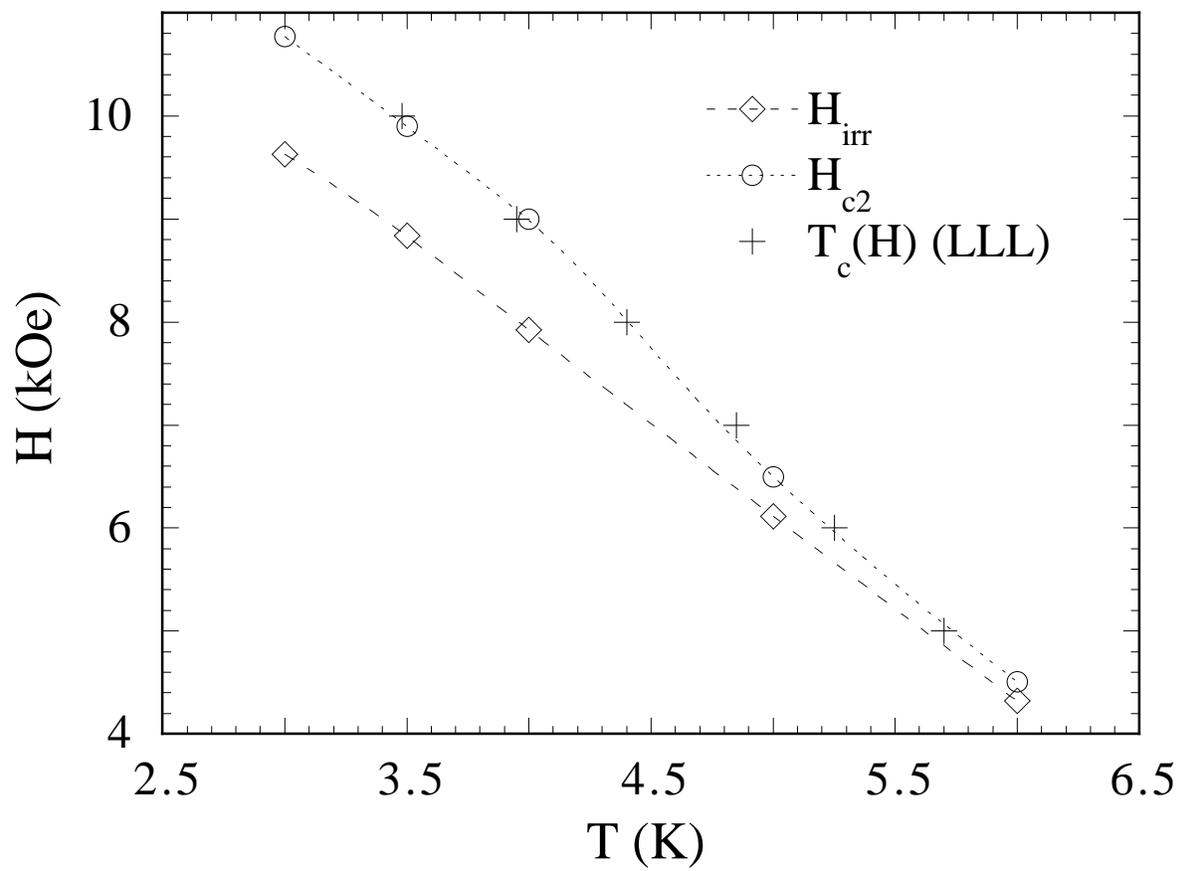
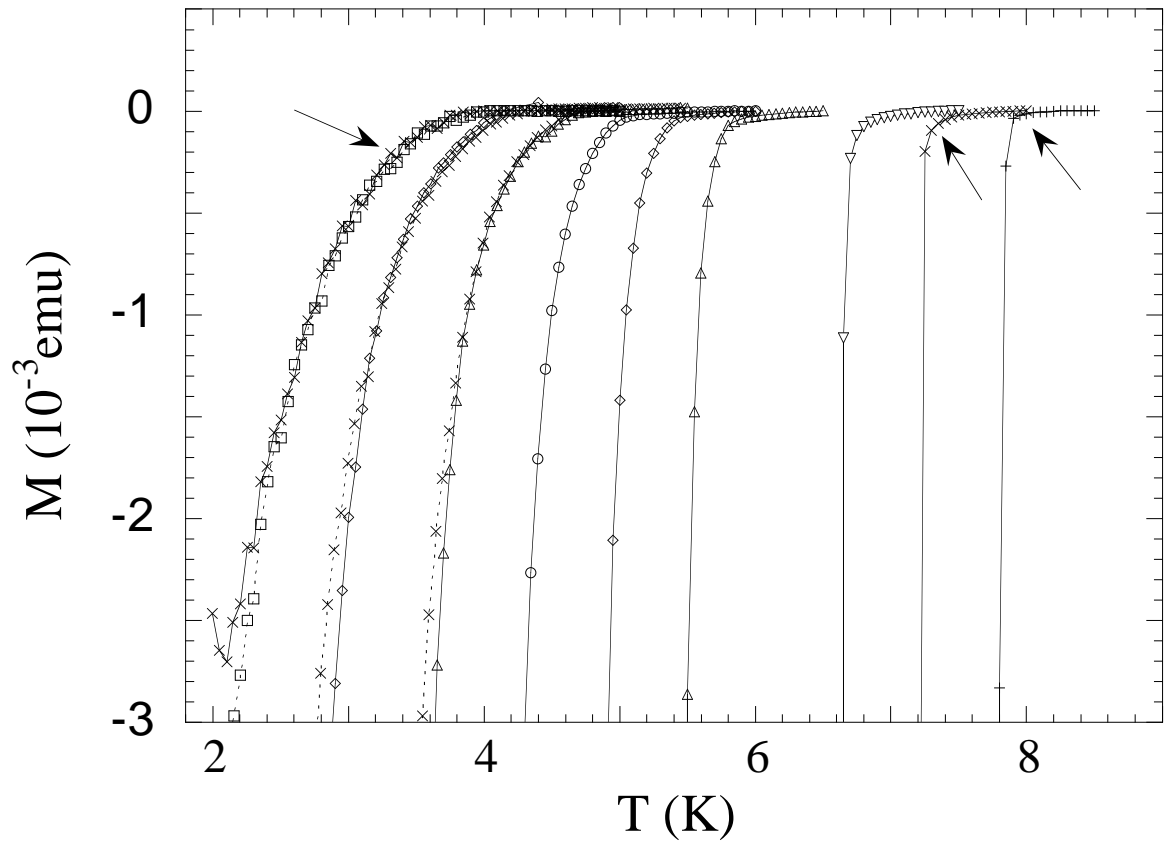
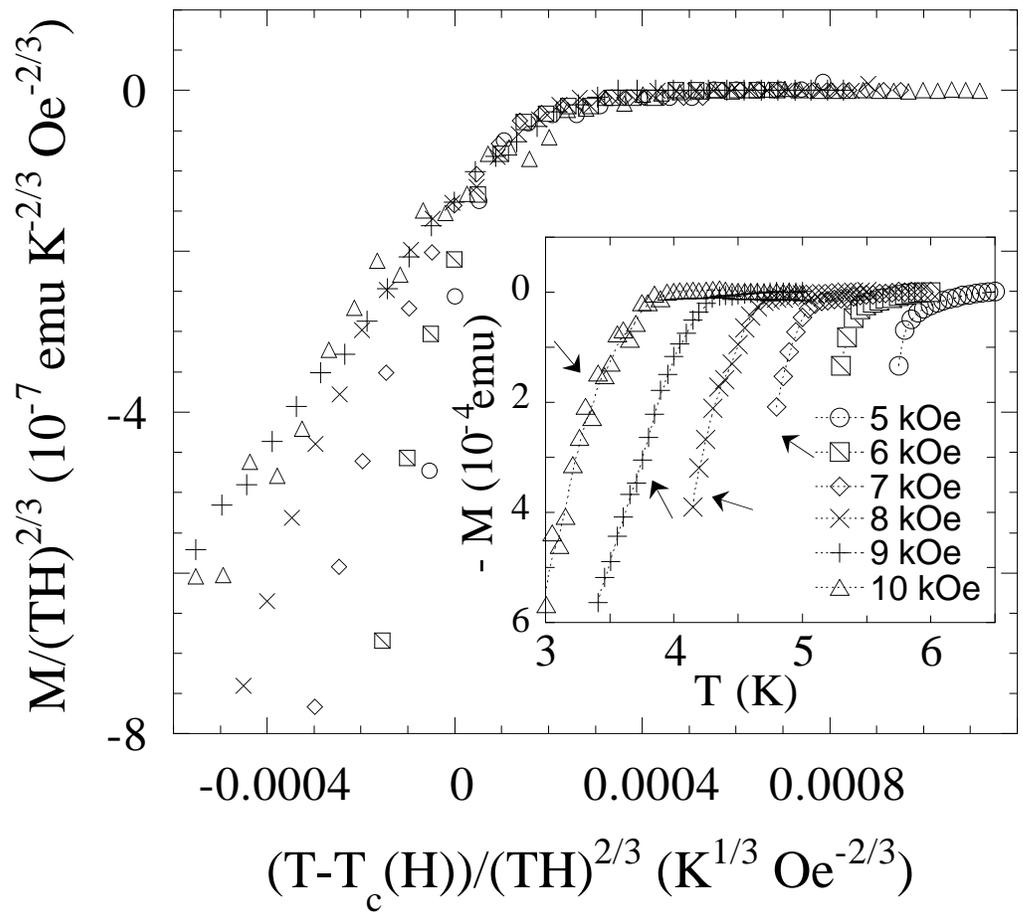
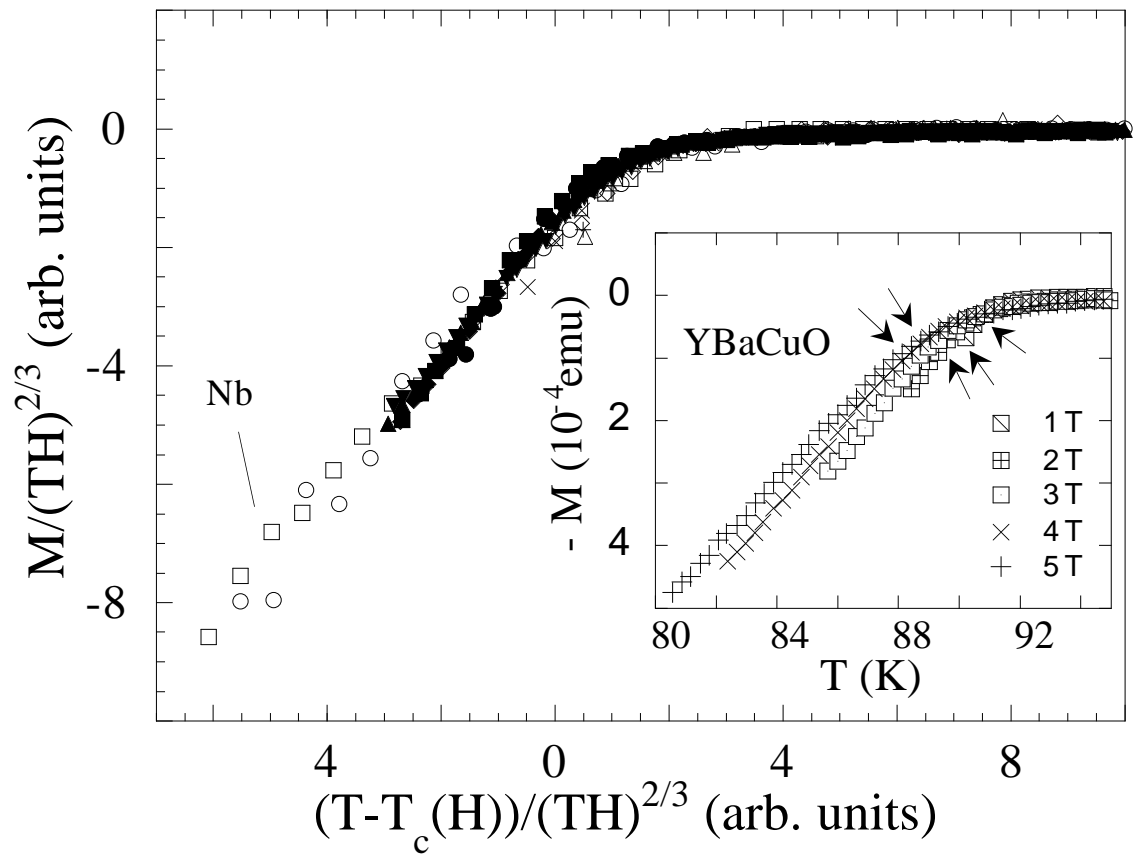
\end{document}